\documentclass[twocolumn,english,prl]{revtex4-1}

\usepackage[T1]{fontenc}
\usepackage[latin9]{inputenc}
\usepackage{color}
\usepackage{babel}
\usepackage{bm}
\usepackage{amsmath}
\usepackage{amsthm}
\usepackage{amssymb}
\usepackage{graphicx}
\usepackage[unicode=true,pdfusetitle,
 bookmarks=true,bookmarksnumbered=false,bookmarksopen=false,
 breaklinks=false,pdfborder={0 0 0},pdfborderstyle={},backref=false,colorlinks=true]
 {hyperref}
\hypersetup{
 urlcolor=blue, citecolor=blue, linkcolor=blue}

\makeatletter
\theoremstyle{remark}
\newtheorem*{acknowledgement*}{\protect\acknowledgementname}

\@ifundefined{definecolor}
 {\usepackage{color}}{}
\makeatother\makeatother

\makeatother

\usepackage{babel}

\makeatother

\usepackage{babel}

\makeatother

\usepackage{babel}

\makeatother

\usepackage{babel}

\makeatother

\usepackage{babel}

\makeatother

\usepackage{babel}

\makeatother

\usepackage{babel}

\makeatother

\usepackage{babel}

\makeatother

\usepackage{babel}

\makeatother

\usepackage{babel}

\makeatother

\usepackage{babel}

\makeatother

\usepackage{babel}

\makeatother

\makeatother

\providecommand{\acknowledgementname}{Acknowledgement}

\begin{document}

\title{Information-disturbance trade-off in generalized entanglement swapping}

\author{Pratapaditya Bej}
\email{pratap6906@gmail.com}

\affiliation{Centre for Astroparticle Physics and Space Science, Bose Institute,
EN 80, Sector V, Bidhannagar, Kolkata 700 091, India}

\author{Arkaprabha Ghosal}
\email{a.ghosal1993@gmail.com}

\affiliation{Centre for Astroparticle Physics and Space Science, Bose Institute,
EN 80, Sector V, Bidhannagar, Kolkata 700 091, India}

\author{Debarshi Das}
\email{dasdebarshi90@gmail.com}

\affiliation{Centre for Astroparticle Physics and Space Science, Bose Institute,
EN 80, Sector V, Bidhannagar, Kolkata 700 091, India}

\author{Arup Roy}
\email{arup145.roy@gmail.com}

\affiliation{S. N. Bose National Center for Basic Sciences, Block JD, Sector III,
Bidhannagar, Kolkata 700098, India}

\author{Somshubhro Bandyopadhyay}
\email{som@jcbose.ac.in}

\affiliation{Department of Physics and Centre for Astroparticle Physics and Space
Science, Bose Institute, EN 80, Sector V, Bidhannagar, Kolkata 700
091, India}
\begin{abstract}
We study information-disturbance trade-off in generalized entanglement
swapping protocols wherein starting from Bell pairs $\left(1,2\right)$
and $\left(3,4\right)$, one performs an arbitrary joint measurement
on $\left(2,3\right)$, so that $\left(1,4\right)$ now becomes correlated.
We obtain trade-off inequalities between information gain in correlations
of $\left(1,4\right)$ and residual information in correlations of
$\left(1,2\right)$ and $\left(3,4\right)$, respectively, and we
argue that information contained in correlations (\emph{information})
is conserved if each inequality is an equality. We show that\emph{
information} is conserved for a maximally entangled measurement but
is not conserved for any other complete orthogonal measurement and
Bell measurement mixed with white noise. However, rather surprisingly,
we find that \emph{information} is conserved for rank-2 Bell diagonal
measurements, although such measurements do not conserve entanglement.
We also show that a separable measurement on $\left(2,3\right)$ can
conserve \emph{information, }even if, as in our example, the post-measurement
states of all three pairs $\left(1,2\right)$, $\left(3,4\right)$,
and $\left(1,4\right)$ become separable. This implies that correlations
from an entangled pair can be transferred to separable pairs in nontrivial
ways so that no \emph{information} is lost in the process. 
\end{abstract}
\maketitle

\section{Introduction}

Quantum systems that have never interacted in the past can nevertheless
become entangled via the procedure of entanglement swapping \citep{swapping}.
The basic idea of entanglement swapping is simple and can be illustrated
with the following protocol: Alice shares a Bell pair $\left(1,2\right)$
with Bob and Bob shares another Bell pair $\left(3,4\right)$ with
Charlie, where all three of them are physically separated; Bob performs
Bell measurement on $\left(2,3\right)$ and discloses the outcome
to both Alice and Charlie, and as Bob's measurement projects $\left(2,3\right)$
onto a Bell state, the result is that Alice and Charlie end up with
a Bell pair $\left(1,4\right)$ as well.

The protocol just described mimics quantum teleportation of a qubit
that is maximally entangled with another qubit, and this is why entanglement
swapping is often considered as teleportation of entanglement. The
protocol, however, can be generalized in several ways, such as modifying
the initial states, Bob's measurement, or both, where some of the
generalizations, as one may note, go beyond that of teleportation
of entanglement. Generalized entanglement swapping protocols have
been well-studied (e.g., \citep{Bose-1998-swapping-multiparty,Gour-Sanders-2004})
and have found important applications in quantum networks \citep{Perseguers-2008-networks,networks-review}
and quantum nonlocality related problems \citep{swapping-CHSH,swapping-activation}.

A generic feature of all ``entanglement swapping'' protocols is
that at the end of a protocol, the pairs $\left(1,2\right)$ and $\left(3,4\right)$
no longer remain as correlated as they were prior to Bob's measurement,
whereas the pair $\left(1,4\right)$ that was completely uncorrelated
earlier becomes correlated. Thus ``entanglement swapping'' can also
be seen as a procedure that, in effect, transfers correlations from
entangled pairs to an unentangled pair, where the transfer of correlations
is brought about by Bob's measurement that disturbs the initial states.
This suggests some kind of information-disturbance trade-off for Bob's
measurements once we interpret ``information'' as information contained
in correlations of a two-qubit state. In this paper, we make this
intuition precise in terms of an information theoretic measure of
entanglement \citep{Zeilinger-99,BZ-1999,BZZ-2001}. This measure,
denoted by $\mathcal{I}\left(\rho\right)$, quantifies the total amount
of information contained in correlations of a two-qubit state $\rho$
(often we shall use \emph{information} as short for ``information
contained in correlations'' when there is no scope for confusion).
Our motivation, in part, comes from an earlier work \citep{complementarity}
on complementarity and information.

Specifically, we consider protocols in which starting from Bell pairs
$\left(1,2\right)$ and $\left(3,4\right)$, Bob performs an arbitrary
two-qubit measurement on $\left(2,3\right)$, so that $\left(1,4\right)$
now becomes correlated. As the initial states are taken to be maximally
entangled, the nonlocal properties of $\left(1,4\right)$ will depend
only on Bob's measurement. This allows us to investigate information-disturbance
type trade-off in a particularly clean set-up. 

Let us now briefly discuss our results. Let $\mathcal{I}_{ij}^{f}$
denote the information contained in correlations of the pair $\left(i,j\right)$
post Bob's measurement. We will show that the following inequalities
hold: 
\begin{eqnarray}
\bar{\mathcal{I}}_{m\,m+1}^{f}+\bar{\mathcal{I}}_{14}^{f} & \leq & 2,\;m=1,3,\label{trade-offs-avg-2}
\end{eqnarray}
where $\bar{\mathcal{I}}_{14}^{f}$ is the average information gain
in correlations of $\left(1,4\right)$, and $\bar{\mathcal{I}}_{m\,m+1}^{f}$
is the average residual information in correlations of the pair $\left(m,m+1\right)$.
Here, the first inequality captures the trade-off for the bipartition
$A\vert BC$ and the second for $C\vert AB$. This is consistent with
the observation that $\left(1,4\right)$, in fact, gains correlations
at the expense of $\left(1,2\right)$ across $A\vert BC$, but at
the expense of $\left(3,4\right)$ across $C\vert AB$. 

Although in this paper we consider only Bell pairs as initial states,
the inequalities \eqref{trade-offs-avg-2} are completely general
and hold for all choices of two-qubit initial states and Bob's measurement
(see Sec.$\,$III for details). 

The physical significance of \eqref{trade-offs-avg-2} can be understood
as follows. For a maximally entangled $\rho$, it holds that $\mathcal{I}\left(\rho\right)=2$
\citep{BZ-1999}, so we can write \eqref{trade-offs-avg-2} as 
\begin{eqnarray}
\bar{\mathcal{I}}_{m\,m+1}^{f}+\bar{\mathcal{I}}_{14}^{f} & \leq & \mathcal{I}_{m\,m+1}^{i},\;m=1,3,\label{trade-offs-avg-2-1}
\end{eqnarray}
where $\mathcal{I}_{m\,m+1}^{i}$ denotes information contained in
correlations of the pair $\left(m,m+1\right)$ prior to Bob's measurement.
Now we have a clear physical interpretation of \eqref{trade-offs-avg-2}:
If equality holds, then no \emph{information} flows out of the system
and the amount of \emph{information }gained in $\left(1,4\right)$
is exactly equal to the amount of \emph{information} lost in $\left(m,m+1\right)$
for $m=1,3$. On the other hand, if any of them is strict then there
is loss of information. Thus, \emph{information} is conserved across
the bipartitions $A\vert BC$ and $C\vert AB$ iff equality holds
in \eqref{trade-offs-avg-2}.

Naturally, the question is, which measurements conserve \emph{information}
in our protocols? It is easy to see \emph{information} is conserved
for maximally entangled measurements, such as Bell measurement. But
are there other measurements with the same property? In particular,
are there measurements that conserve \emph{information} but not entanglement?
Intuitively, it appears that only maximally entangled measurements
would conserve \emph{information}, and other measurements would necessarily
lead to \emph{information} loss. We find that this is indeed the case
for all complete orthogonal (but not maximally entangled) measurements
and Bell measurement mixed with white noise. But surprisingly, it
turns out there are exceptions: in particular, rank-2 Bell-diagonal
measurements conserve \emph{information} over all intermediate strengths.
Interestingly, \emph{information} is conserved even when they are
separable, and all post-measurement states become separable after
Bob's measurement. The latter suggests that separable measurements
can distribute correlations from an entangled pair to separable pairs
in ways such that \emph{information} is conserved. 

Here we want to emphasize that we are able to express \eqref{trade-offs-avg-2}
in the form of \eqref{trade-offs-avg-2-1} only because the initial
states are maximally entangled. In this case, we can infer that \emph{information}
can never be increased post Bob's measurement and classical communication
of the measurement outcome. However, if the initial states are not
maximally entangled, one has $\mathcal{I}_{m\,m+1}^{i}<2$ for $m=1,3$,
and in such cases \eqref{trade-offs-avg-2-1} does not follow from
\eqref{trade-offs-avg-2} because $\mathcal{I}$ is not a LOCC (Local
Operations and Classical Communication) monotone \citep{I-not-monotone}.
However, as noted earlier, the inequalities \eqref{trade-offs-avg-2}
must always hold in all entanglement swapping protocols, hence they
are of basic importance.

The paper is organized as follows. In Sec.$\,$II, we discuss the
basics of the entanglement swapping protocol considered in this paper.
Here we show that the post-measurement state of $\left(1,4\right)$
for any given outcome is completely specified in terms of the corresponding
POVM element; however, similar expressions for $\left(1,2\right)$
and $\left(3,4\right)$ could not be obtained for they do not seem
to have any simplified form in general, though it is clear they depend
only on the POVM elements, as expected. Next we consider a general
family of Bell-diagonal measurements and obtain the post-measurement
states for all three pairs $\left(1,4\right)$, $\left(1,2\right)$,
and $\left(3,4\right)$. In Sec.$\,$III, we obtain the trade-off
relations \eqref{trade-offs-avg-2}. We study the trade-off relations
for complete orthogonal measurements and Bell-diagonal measurements
in Sec.$\,$IV. We conclude with a brief summary of the paper and
a short discussion on open questions in Sec.$\,$V. 

\section{generalized protocol}

In this section, we first study how the initial Bell states transform
under Bob's two-qubit measurement $\bm{\Pi}$ (a POVM) and obtain
an expression for the post-measurement state of $\left(1,4\right)$.
After that, we consider a family of Bell-diagonal measurements and
obtain expressions for the post-measurement states for all three pairs.
These expressions will be used to analyze the trade-off relations
for specific measurements in Section IV.

To begin with, Alice and Charlie each shares a Bell state $\left|\Psi_{1}\right\rangle $
with Bob, where $\left|\Psi_{1}\right\rangle $ belongs to the two-qubit
Bell basis 
\begin{eqnarray}
\left|\Psi_{1}\right\rangle =\frac{1}{\sqrt{2}}\left(\left|00\right\rangle +\left|11\right\rangle \right); &  & \left|\Psi_{2}\right\rangle =\frac{1}{\sqrt{2}}\left(\left|00\right\rangle -\left|11\right\rangle \right);\nonumber \\
\left|\Psi_{3}\right\rangle =\frac{1}{\sqrt{2}}\left(\left|01\right\rangle +\left|10\right\rangle \right); &  & \left|\Psi_{4}\right\rangle =\frac{1}{\sqrt{2}}\left(\left|01\right\rangle -\left|10\right\rangle \right).\label{Bell-basis}
\end{eqnarray}
 Then, the four-qubit initial state is given by 
\begin{eqnarray}
\left|\Phi\right\rangle  & = & \left|\Psi_{1}\right\rangle _{12}\left|\Psi_{1}\right\rangle _{34},\label{initial state}
\end{eqnarray}
where$\left(1,2\right)$ is shared between Alice and Bob, and $\left(3,4\right)$
is shared between Bob and Charlie. 

Bob's measurement $\bm{\Pi}$ can be specified by a collection of
positive semi-definite operators $\Pi_{i}$ satisfying $\sum\Pi_{i}=\mathbb{I}$,
where each POVM element admits decomposition of the form 
\begin{eqnarray}
\Pi_{k} & = & \sum_{\alpha=1}^{4}\pi_{k\alpha}\left|\phi_{k\alpha}\right\rangle \left\langle \phi_{k\alpha}\right|,\label{Pi-spectral}
\end{eqnarray}
where $\left|\phi_{k\alpha}\right\rangle $, $\alpha=1,\dots,4$ form
a complete orthonormal basis on $\mathbb{C}^{2}\otimes\mathbb{C}^{2}$,
$\pi_{k\alpha}\geq0$ for $\alpha=1,\dots,4$. Note that, if $\Pi_{k}$
is not full-rank, the decomposition \eqref{Pi-spectral} is not unique,
but as we will see, this will not affect post-measurement states.

The probability of obtaining outcome $k$ is 
\begin{eqnarray}
p_{k} & = & \left\langle \Phi\left|\Pi_{k}\otimes\mathbb{I}\right|\Phi\right\rangle =\frac{1}{4}\mbox{Tr}\Pi_{k}\label{probability-kth-outcome}
\end{eqnarray}
and the post-measurement four-qubit state is given by 
\begin{eqnarray}
\left|\Phi_{k}\right\rangle  & = & \frac{1}{\sqrt{p_{k}}}\sqrt{\Pi_{k}}\otimes\mathbb{I}\left|\Phi\right\rangle ,\label{post-measurement-Phi(k)}
\end{eqnarray}
where it is understood that $\Pi_{k}$ acts on the pair $\left(2,3\right)$
and the identity operator acts on the pair $\left(1,4\right)$.

To find an expression for \eqref{post-measurement-Phi(k)} we proceed
as follows. First we write the initial state \eqref{initial state}
as 
\begin{eqnarray}
\left|\Phi\right\rangle  & = & \frac{1}{2}\sum_{i=1}^{4}\left|\Psi_{i}\right\rangle _{23}\left|\Psi_{i}\right\rangle _{14}.\label{initial-2}
\end{eqnarray}

We now use the fact that \eqref{initial-2} is $U\otimes U^{*}$ invariant
and therefore can be written as 
\begin{eqnarray}
\left|\Phi\right\rangle  & = & \frac{1}{2}\sum_{\alpha=1}^{4}\left|\phi_{k\alpha}\right\rangle _{23}\left|\phi_{k\alpha}^{*}\right\rangle _{14},\label{initial-2-1}
\end{eqnarray}
where the complex conjugation is in the computational basis. Then,
using \eqref{initial-2-1} and \eqref{Pi-spectral} in \eqref{post-measurement-Phi(k)}
we get 
\begin{eqnarray}
\left|\Phi_{k}\right\rangle  & = & \frac{1}{2\sqrt{p_{k}}}\sum_{\alpha=1}^{4}\sqrt{\pi_{k\alpha}}\left|\phi_{k\alpha}\right\rangle _{23}\left|\phi_{k\alpha}^{*}\right\rangle _{14}.\label{phi_k-general}
\end{eqnarray}
The above expression will be used to find the post-measurement states. 

By tracing out qubits 2 and 3 from \eqref{phi_k-general} we immediately
obtain the post-measurement state of $\left(1,4\right)$: 
\begin{eqnarray}
\rho_{14\vert k} & = & \frac{1}{4p_{k}}\sum_{\alpha=1}^{4}\pi_{k\alpha}\left|\phi_{k\alpha}^{*}\right\rangle \left\langle \phi_{k\alpha}^{*}\right|=\frac{\Pi_{k}^{*}}{\mbox{Tr}\left(\Pi_{k}\right)},\label{rho14|k}
\end{eqnarray}
where $\Pi_{k}^{*}=\sum_{\alpha=1}^{4}\pi_{k,\alpha}\left|\phi_{k\alpha}^{*}\right\rangle \left\langle \phi_{k\alpha}^{*}\right|$. 

Obtaining similar compact expressions for $\left(1,2\right)$ and
$\left(3,4\right)$, however, seems difficult, unless we know more
about the measurement itself. So now we turn our attention to Bell-diagonal
measurements and show that it is in fact possible to obtain exact
expressions for all post-measurement states.

\subsection*{Bell-diagonal measurements}

We consider a family of Bell-diagonal measurements $\mathbb{M}\left(\lambda\right)$
characterized by a real variable $\lambda$ ($0\leq\lambda\leq1$).
The POVM elements $\mathbb{M}_{i}\left(\lambda\right)$, $i=1,\dots,4$
are positive semi-definite operators satisfying $\sum_{i}\mathbb{M}_{i}\left(\lambda\right)=\mathbb{I}$
and are given by 
\begin{eqnarray}
\mathbb{M}_{i}\left(\lambda\right) & = & \lambda\left|\Psi_{i}\right\rangle \left\langle \Psi_{i}\right|+\left(1-\lambda\right)\sum_{l=1}^{4}q_{il}\left|\Psi_{l}\right\rangle \left\langle \Psi_{l}\right|,\label{M(i)-lambda}
\end{eqnarray}
 where for every $i$, $0\leq q_{il}\leq1$ and $\sum_{i=1}^{4}q_{il}=1$
for every $l=1,\dots,4$. The POVM elements can be explicitly written
in the Bell-diagonal form 
\begin{eqnarray}
\mathbb{M}_{i}\left(\lambda\right) & = & x_{i}\left|\Psi_{i}\right\rangle \left\langle \Psi_{i}\right|+\sum_{j\neq i}y_{ij}\left|\Psi_{j}\right\rangle \left\langle \Psi_{j}\right|\label{eM_i-lambda}
\end{eqnarray}
 for $i,j=1,\dots,4$, where $x_{i}=\left[\lambda+\left(1-\lambda\right)q_{ii}\right]$
and $y_{ij}=\left(1-\lambda\right)q_{ij}$.

The probability of outcome $k$ is obtained from \eqref{probability-kth-outcome}:

\begin{eqnarray}
p_{k} & = & \frac{1}{4}\left(x_{k}+\sum_{j\neq k}y_{kj}\right).\label{p(k)-Bell-diagonal}
\end{eqnarray}
 The post-measurement state of $\left(1,4\right)$ is obtained from
\eqref{rho14|k}: 
\begin{eqnarray}
\rho_{14\vert k} & = & \frac{1}{x_{k}+\underset{j\neq k}{\sum}y_{kj}}\mathbb{M}_{k}\left(\lambda\right)\label{rh_14-unsharp}
\end{eqnarray}
 Now we need to find $\rho_{12\vert k}$ and $\rho_{34\vert k}$.
First we observe that the following identities hold for a four qubit-system:
\begin{eqnarray*}
\left|\Psi_{1}\right\rangle _{12}\left|\Psi_{1}\right\rangle _{34} & = & \frac{1}{2}\sum_{i=1}^{4}\left|\Psi_{i}\right\rangle _{23}\left|\Psi_{i}\right\rangle _{14},\\
\left|\Psi_{2}\right\rangle _{12}\left|\Psi_{2}\right\rangle _{34} & = & \frac{1}{2}\sum_{i=1,2}\left|\Psi_{i}\right\rangle _{23}\left|\Psi_{i}\right\rangle _{14}-\frac{1}{2}\sum_{i=3,4}\left|\Psi_{i}\right\rangle _{23}\left|\Psi_{i}\right\rangle _{14},\\
\left|\Psi_{3}\right\rangle _{12}\left|\Psi_{3}\right\rangle _{34} & = & \frac{1}{2}\sum_{i=1,3}\left|\Psi_{i}\right\rangle _{23}\left|\Psi_{i}\right\rangle _{14}-\frac{1}{2}\sum_{i=2,4}\left|\Psi_{i}\right\rangle _{23}\left|\Psi_{i}\right\rangle _{14},\\
\left|\Psi_{4}\right\rangle _{12}\left|\Psi_{4}\right\rangle _{34} & = & \frac{1}{2}\sum_{i=1,4}\left|\Psi_{i}\right\rangle _{23}\left|\Psi_{i}\right\rangle _{14}-\frac{1}{2}\sum_{i=2,3}\left|\Psi_{i}\right\rangle _{23}\left|\Psi_{i}\right\rangle _{14}.
\end{eqnarray*}
The above identities can be written in a compact form as 
\begin{eqnarray}
\left|\Psi_{i}\right\rangle _{12}\left|\Psi_{i}\right\rangle _{34} & = & \frac{1}{2}\sum_{l=1}^{4}\mu_{il}\left|\Psi_{l}\right\rangle _{23}\left|\Psi_{l}\right\rangle _{14},\label{identity}
\end{eqnarray}
where $i=1,\dots,4$, and $\mu_{il}\in\left\{ +1,-1\right\} $. Now
substituting \eqref{eM_i-lambda} in \eqref{post-measurement-Phi(k)}
and after simplification using \eqref{identity}, we find that 
\begin{eqnarray}
\left|\Phi_{k}\right\rangle  & = & \frac{1}{4\sqrt{p_{k}}}\sum_{l=1}^{4}\gamma_{l}\left|\Psi_{l}\right\rangle _{12}\left|\Psi_{l}\right\rangle _{34},
\end{eqnarray}
where $\gamma_{l}=\left(\sqrt{x_{k}}\mu_{kl}+\sum_{j\neq k}\sqrt{y_{kj}}\mu_{jl}\right)$
and $p_{k}$ is given by \eqref{p(k)-Bell-diagonal}. Then the post-measurement
states $\rho_{12\vert k}$ and $\rho_{34\vert k}$ are given by 
\begin{eqnarray}
\rho_{m\,m+1\vert k} & = & \sum_{l=1}^{4}\tau_{l}\left|\Psi_{l}\right\rangle \left\langle \Psi_{l}\right|,\;m=1,3,\label{rho_12|kandrho_34|k}
\end{eqnarray}
where $\tau_{l}=\frac{\gamma_{l}^{2}}{16p_{k}}$ for $l=1,\dots,4$.
So all post-measurement states are Bell-diagonal.

\section{Trade-off relations }

We begin with the definition of $\mathcal{I}\left(\rho\right)$ --
the total information contained in correlations of a two-qubit state
$\rho$. It is defined as \citep{BZ-1999,BZZ-2001} 
\begin{eqnarray}
\mathcal{I}\left(\rho\right) & = & \max_{\bm{n},\bm{m}}\left(I_{nn}+I_{mm}\right),\label{I-definition}
\end{eqnarray}
where $I_{kk}=\left[\mbox{Tr}\rho\left(\sigma_{k}\otimes\sigma_{k}\right)\right]^{2}$
for a unit vector $\bm{k}\in\mathbb{R}^{3}$, $\sigma_{n}=\bm{\sigma\cdot n}$,
$\sigma_{x}.\sigma_{y},\sigma_{z}$ being the standard Pauli matrices,
and the maximum is taken over all pairs $\left(\bm{n},\bm{m}\right)$
of unit vectors with the property that the spin measurements along
$\bm{n}$ and along $\bm{m}$ are mutually unbiased.

The following properties hold: 
\begin{itemize}
\item $\mathcal{I}\left(\rho\right)\leq2$, where the equality is achieved
for maximally entangled states; 
\item $\mathcal{I}\left(\rho\right)\leq1$ for separable states; 
\item $\mathcal{I}\left(\rho\right)=\frac{\mathcal{B}^{2}\left(\rho\right)}{4}$
\citep{BZZ-2001,complementarity}, where $\mathcal{B}\left(\rho\right)$
is the maximal mean value of the Bell-CHSH observable \citep{Horodecki-1995-Bell,Horodecki-1996-Bell-2}. 
\end{itemize}
The last property is particularly important for two reasons. First,
it provides us with a computable formula for $\mathcal{I}\left(\rho\right)$
(see next section) and second, it plays a crucial role in the derivation
of the trade-off relations. Note however that, though $\mathcal{I}\left(\rho\right)$
and $\mathcal{B}\left(\rho\right)$ are related for any two-qubit
state, such a relation may not hold in higher dimensions, and moreover,
$\mathcal{I}\left(\rho\right)$ and Bell-CHSH violation are conceptually
inequivalent.

The trade-off relations are obtained by exploiting the monogamy property
of $\mathcal{I}$. This property follows from the Bell-monogamy \citep{Bell-monogamy,Bell-monogamy-2}:
For any three qubit state $\sigma_{abc}$ it holds that 
\begin{eqnarray}
\mathcal{B}^{2}\left(\sigma_{ab}\right)+\mathcal{B}^{2}\left(\sigma_{ac}\right) & \leq & 8,\label{Bell-monogamy}
\end{eqnarray}
where $\sigma_{ab}=\mbox{Tr}_{c}\left(\sigma_{abc}\right)$, $\sigma_{ac}=\mbox{Tr}_{b}\left(\sigma_{abc}\right)$.
Similar inequalities hold for other possible pairs. 

Since $\mathcal{I}\left(\rho\right)=\frac{\mathcal{B}^{2}\left(\rho\right)}{4}$,
\eqref{Bell-monogamy} therefore leads to the $\mathcal{I}$-monogamy
inequality 
\begin{eqnarray}
\mathcal{I}\left(\sigma_{ab}\right)+\mathcal{I}\left(\sigma_{ac}\right) & \leq & 2\label{I-abc-<=00003D00003D2}
\end{eqnarray}
for any three qubit state $\sigma_{abc}$.

To obtain the trade-off relations \eqref{trade-offs-avg-2} we now
proceed as follows. Let $\rho_{1234}$ be the four-qubit state post
Bob's measurement and $\rho_{ij}$ be the post-measurement states
of the pairs $\left(i,j\right)$ obtained from $\rho_{1234}$ by tracing
out the appropriate qubits, e.g. $\rho_{14}=\text{Tr}_{23}\left(\rho_{1234}\right)$
and so on. Then applying \eqref{I-abc-<=00003D00003D2} to each of
the three-qubit states $\rho_{124}=\mbox{Tr}_{3}\left(\rho_{1234}\right)$
and $\rho_{134}=\mbox{Tr}_{2}\left(\rho_{1234}\right)$ we get the
following inequalities 
\begin{eqnarray}
\mathcal{I}_{m\,m+1}^{f}+\mathcal{I}_{14}^{f} & \leq & 2,\;m=1,3\label{trade-offs}
\end{eqnarray}
that hold for any outcome of Bob's measurement. Here, $\mathcal{I}_{14}^{f}$
amounts to the gain in information in correlations of $\left(1,4\right)$
as $\mathcal{I}_{14}^{i}=0$ initially, whereas $\mathcal{I}_{m\,m+1}^{f}$
amounts to the residual information contained in correlations of $\left(m,m+1\right)$
for $m=1,3$. 

Now, for any given measurement there are three possible scenarios:
the inequalities \eqref{trade-offs} are strict for all outcomes,
some outcomes, or none of the outcomes. So one must consider weighted
average of \eqref{trade-offs} taken over all possible outcomes and
that gives us the desired relations \eqref{trade-offs-avg-2} (reproduced
here for convenience) 
\begin{eqnarray}
\bar{\mathcal{I}}_{m\,m+1}^{f}+\bar{\mathcal{I}}_{14}^{f} & \leq & 2,\;m=1,3\label{trade-offs-avg}
\end{eqnarray}
where $\overline{\mathcal{I}}$ denotes the weighted average over
all outcomes. 

As we explained in the introduction, \emph{information} is conserved
across the bipartitions provided equality holds in each of the above
inequalities. So now we want to find out which measurements conserve
\emph{information}. In the next section, we present partial answers
and some useful insights. 

\section{trade-off relations for specific measurements}

To study the trade-off relations we need to be able to compute $\mathcal{I\left(\rho\right)}$
for any two-qubit $\rho$. Fortunately, it is possible by virtue of
the relation $\mathcal{I}\left(\rho\right)=\frac{\mathcal{B}^{2}\left(\rho\right)}{4}$. 

Let $t_{11},t_{22},t_{33}$ be the eigenvalues of the real $3\times3$
matrix $T$, the elements of which are defined as $T_{ij}={\rm Tr}\left[\rho\left(\sigma_{i}\otimes\sigma_{j}\right)\right]$
for $i,j=1,2,3$. Define the function 
\begin{eqnarray}
M\left(\rho\right) & = & \max_{i>j}\left(\left|t_{ii}\right|^{2}+\left|t_{jj}\right|^{2}\right).\label{M(rho)}
\end{eqnarray}

Then, $\mathcal{B}\left(\rho\right)=2\sqrt{M\left(\rho\right)}$ \citep{Horodecki-1995-Bell}
and consequently, 
\begin{eqnarray}
\mathcal{I}\left(\rho\right) & = & M\left(\rho\right).\label{I(rho)=00003DM(rho)}
\end{eqnarray}

\subsection*{Complete orthogonal measurements}

In a complete orthogonal measurement (COM) $\left\{ \Pi_{k}\right\} $,
each POVM element $\Pi_{k}$ is a projection operator $\left|\eta_{k}\right\rangle \left\langle \eta_{k}\right|$
that projects the system onto $\left|\eta_{k}\right\rangle $, which
is an eigenvector of the measurement, and these eigenvectors $\left\{ \left|\eta_{k}\right\rangle \right\} $
form a complete orthonormal basis of the Hilbert space of the system,
i.e., $\text{Tr}\Pi_{k}\Pi_{l}=\delta_{kl}$. The simplest example
of a COM is the Bell measurement for which \eqref{trade-offs-avg}
are in fact equalities, so \emph{information} is conserved.

Now consider a COM, which is not maximally entangled. Then at least
one of the eigenvectors must be nonmaximally entangled. Let this eigenvector
be $\left|\phi\right\rangle $. When Bob's measurement projects $\left(2,3\right)$
onto $\left|\phi\right\rangle $ (which happens with some nonzero
probability {[}see, \eqref{probability-kth-outcome}{]}), the pair
$\left(1,4\right)$ collapses onto $\left|\phi^{*}\right\rangle $,
where complex conjugation is taken with respect to the computational
basis. Note that, as $\left|\phi\right\rangle $ is nonmaximally entangled,
so is $\left|\phi^{*}\right\rangle $; in fact, complex conjugation,
although a nonlocal operation, does not change entanglement properties
of pure states. 

Let $\alpha,\beta$ be the Schmidt coefficients of $\left|\phi^{*}\right\rangle $,
where without loss of generality we assume that $\alpha>\beta$. A
simple calculation using \eqref{I(rho)=00003DM(rho)} shows that 
\begin{eqnarray*}
\mathcal{I}_{14}^{f} & = & 1+4\alpha^{2}\beta^{2},\\
\mathcal{I}_{m\,m+1}^{f} & = & \left(\alpha^{2}-\beta^{2}\right)^{4},\;m=1,3.
\end{eqnarray*}
The trade-off inequalities \eqref{trade-offs} for this particular
outcome are therefore given by 
\begin{eqnarray}
\text{\ensuremath{\mathcal{I}_{m\,m+1}^{f}+\mathcal{I}_{14}^{f}}}=\left(\alpha^{2}-\beta^{2}\right)^{4}+1+4\alpha^{2}\beta^{2},m=1,3. & \text{\;}\label{trade-off-COM}
\end{eqnarray}
Now, $\left(\alpha^{2}-\beta^{2}\right)^{4}<\left(\alpha^{2}-\beta^{2}\right)^{2}$
as $0<\left(\alpha^{2}-\beta^{2}\right)<1$, and therefore we can
write \eqref{trade-off-COM} as 
\begin{eqnarray*}
\text{\ensuremath{\mathcal{I}_{m\,m+1}^{f}+\mathcal{I}_{14}^{f}}} & < & \left(\alpha^{2}-\beta^{2}\right)^{2}+1+4\alpha^{2}\beta^{2}=2;\;m=1,3.
\end{eqnarray*}
Thus the inequalities \eqref{trade-off-COM} are strict for this outcome;
hence, \eqref{trade-offs-avg} are strict as well. So we conclude
that \emph{information} is not conserved for COMs that are not maximally
entangled.

\begin{figure}
\includegraphics[scale=0.4]{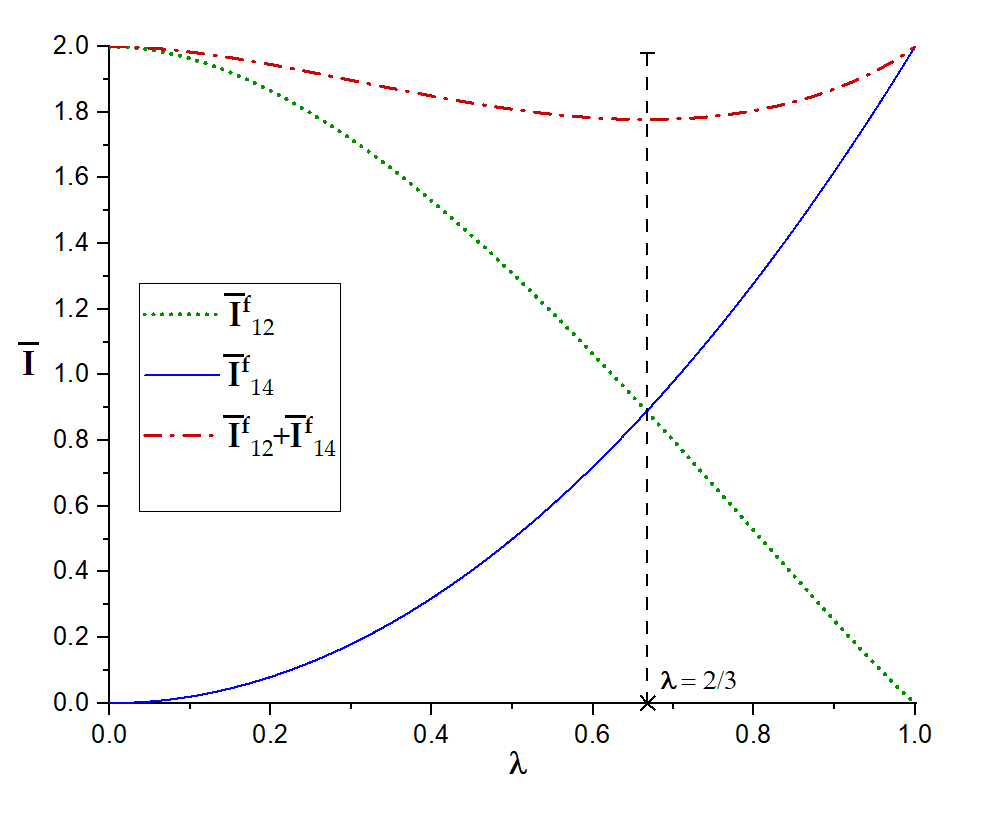}\caption{Bell measurement mixed with white noise: Behaviour of $\mathcal{I}_{14}^{f}$
, $\mathcal{I}_{12}^{f}$, and $\mathcal{I}_{{\rm tot}}^{f}=\mathcal{I}_{12}^{f}+\mathcal{I}_{14}^{f}$
as function of $\lambda\in\left[0,1\right]$. }
\end{figure}

\subsection*{Bell measurement mixed with white noise}

The POVM elements are defined as 
\begin{eqnarray*}
\mathbb{M}_{i}\left(\lambda\right) & = & \lambda\left|\Psi_{i}\right\rangle \left\langle \Psi_{i}\right|+\left(1-\lambda\right)\frac{\mathbb{I}}{4}
\end{eqnarray*}
for $i=1,\dots,4$. The measurement is separable for $0\leq\lambda\leq\frac{1}{3}$
and entangled for $\frac{1}{3}<\lambda\leq1$. Comparing with \eqref{eM_i-lambda}
we see that $q_{il}=\frac{1}{4}$ $\forall$ $i,l=1,\dots,4$; $x_{i}=\frac{3\lambda+1}{4}$
and $y_{ij}=\frac{1-\lambda}{4}$ $\forall$ $i,j=1,\dots,4$. Then
from \eqref{p(k)-Bell-diagonal} we find that $p_{k}=\frac{1}{4}$,
so all outcomes are equally probable. The post-measurement states
are obtained from \eqref{rh_14-unsharp} and \eqref{rho_12|kandrho_34|k}:
\begin{eqnarray}
\rho_{14\vert k} & = & a\left|\Psi_{k}\right\rangle \left\langle \Psi_{k}\right|+b\sum_{j\neq k}\left|\Psi_{j}\right\rangle \left\langle \Psi_{j}\right|,\label{rho14|k-white-noise}\\
\rho_{m\,m+1\vert k} & = & c\left|\Psi_{k}\right\rangle \left\langle \Psi_{k}\right|+d\sum_{l\neq k}\left|\Psi_{l}\right\rangle \left\langle \Psi_{l}\right|,\label{rho12|k-whitenoise}
\end{eqnarray}
for $m=1,3$, where
\begin{eqnarray*}
a & = & \frac{3\lambda+1}{4},\\
b & = & \frac{1-\lambda}{4},\\
c & = & \frac{1}{16}\left(\sqrt{3\lambda+1}+3\sqrt{1-\lambda}\right)^{2},\\
d & = & \frac{1}{16}\left(\sqrt{3\lambda+1}-\sqrt{1-\lambda}\right)^{2}.
\end{eqnarray*}
Note that the states $\rho_{14\vert k}$ and $\rho_{14\vert k^{\prime}}$
are LU equivalent for $k,k^{\prime}=1,\dots,4$ and so are $\rho_{m\,m+1\vert k}$
and $\rho_{m\,m+1\vert k^{\prime}}$ for $m=1,3$ and $k,k^{\prime}=1,\dots,4$.
This means that $\mathcal{I}_{14}$, $\mathcal{I}_{12}$, and $\mathcal{I}_{34}$
do not change with outcomes and the trade-off relations \eqref{trade-offs}
will be the same for every outcome. In what follows, we therefore
omit the subscript for measurement outcome.

From \eqref{rho14|k-white-noise} and \eqref{rho12|k-whitenoise}
and using \eqref{I(rho)=00003DM(rho)} we find that 
\begin{eqnarray*}
\mathcal{I}_{14}^{f} & = & 2\lambda^{2},\\
\mathcal{I}_{m\,m+1}^{f} & = & \frac{1}{2}\left(1-\lambda+\sqrt{\left(1-\lambda\right)\left(1+3\lambda\right)}\right)^{2},
\end{eqnarray*}
where $m=1,3$ and $0\leq\lambda\leq1$. Substituting the above in
\eqref{trade-offs} it is easy to show that the inequalities are strict
for all $0<\lambda<1$ (note that, $\lambda=1$ corresponds to the
Bell measurement). So \emph{information} is not conserved for $\lambda\in\left(0,1\right)$.

Let us now look at the process of \emph{information} transfer as we
vary $\lambda$ which controls the strength of the measurement. In
Fig.$\,$1 we have plotted the quantities $\mathcal{I}_{14}^{f}$,
$\mathcal{I}_{12}^{f}$, and $\mathcal{I}_{{\rm tot}}^{f}=\mathcal{I}_{m\,m+1}^{f}+\mathcal{I}_{14}^{f}$,
each of which is a function of $\lambda$. 

Observations: 
\begin{itemize}
\item At $\lambda=0$, $\rho_{14}$ is maximally mixed and $\rho_{12}$,
$\rho_{34}$ are both maximally entangled. So $\mathcal{I}_{{\rm tot}}=2$. 
\item As $\lambda$ starts moving away from zero, we start observing \emph{information}
transfer throughout, even though the measurement remains separable
through $\lambda\in\left[0,\frac{1}{3}\right]$.
\item $\mathcal{I}_{{\rm tot}}<2$ for $\lambda\in\left(0,1\right)$ which
indicates loss of \emph{information} in this range. The \emph{information}
loss is maximum at $\lambda=\frac{2}{3}$ where $\mathcal{I}_{\text{tot}}$
attains the minimum. But note that from this point onward $\mathcal{I}_{\text{tot}}$
starts to increase, eventually reaching the maximum at $\lambda=1$
which corresponds to the Bell measurement. 
\item The crossover between $\mathcal{I}_{14}^{f}$ and $\mathcal{I}_{12}^{f}$
at $\lambda=\frac{2}{3}$ tells us that when the measurement is weak
in the beginning, the initial pairs lose more \emph{information} than
that gained by $\left(1,4\right)$, but as the measurement picks up
strength, this reverses, which is why the minimum is observed.
\end{itemize}
So far we have seen \emph{information} is not conserved for COMs (not
maximally entangled) and for measurements obtained by mixing Bell
measurement with white noise. This prompted us to question whether
this could be a generic feature for all measurements that are not
maximally entangled. But because a general result is quite difficult
to get, we explored the question within the family of Bell-diagonal
measurements $\mathbb{M}\left(\lambda\right)$ and found that there
exists a rank-two measurement for which this is not the case. This
is discussed next.

\subsection*{Rank-two Bell-diagonal measurement }

Here the POVM elements are defined as 
\begin{eqnarray*}
\mathbb{M}_{1}\left(\lambda\right) & = & \lambda\left|\Psi_{1}\right\rangle \left\langle \Psi_{1}\right|+\left(1-\lambda\right)\left|\Psi_{2}\right\rangle \left\langle \Psi_{2}\right|,\\
\mathbb{M}_{2}\left(\lambda\right) & = & \lambda\left|\Psi_{2}\right\rangle \left\langle \Psi_{2}\right|+\left(1-\lambda\right)\left|\Psi_{1}\right\rangle \left\langle \Psi_{1}\right|,\\
\mathbb{M}_{3}\left(\lambda\right) & = & \lambda\left|\Psi_{3}\right\rangle \left\langle \Psi_{3}\right|+\left(1-\lambda\right)\left|\Psi_{4}\right\rangle \left\langle \Psi_{4}\right|,\\
\mathbb{M}_{4}\left(\lambda\right) & = & \lambda\left|\Psi_{4}\right\rangle \left\langle \Psi_{4}\right|+\left(1-\lambda\right)\left|\Psi_{3}\right\rangle \left\langle \Psi_{3}\right|,
\end{eqnarray*}
for $\lambda\in\left[0,1\right]$. The measurement is entangled except
at $\lambda\neq\frac{1}{2}$.

For outcome $k$, the post-measurement states are obtained from \eqref{rh_14-unsharp}
and \eqref{rho_12|kandrho_34|k}; in particular, 
\begin{eqnarray}
\rho_{14\vert k} & = & \mathbb{M}_{k}\left(\lambda\right),\label{rho14-rank-2}\\
\rho_{m\,m+1\vert k} & = & \sum_{l=1}^{4}\tau_{l}\left|\Psi_{l}\right\rangle \left\langle \Psi_{l}\right|,\;m=1,3,\label{rho12/34-rank2}
\end{eqnarray}
where 
\begin{eqnarray*}
\tau_{l} & = & \frac{1}{4}\left(1+2\sqrt{\lambda\left(1-\lambda\right)}\right),\;l=\begin{array}{ccc}
k,k+1 & \mbox{for} & k=1,3\\
k,k-1 & \mbox{for} & k=2,4
\end{array}\\
\tau_{l} & = & \frac{1}{4}\left(1-2\sqrt{\lambda\left(1-\lambda\right)}\right),\;\mbox{ }l\,\mbox{otherwise}.
\end{eqnarray*}
Note that, $\rho_{14\vert k}$ is entangled $\forall\lambda$ except
when $\lambda=\frac{1}{2}$ but $\rho_{12\vert k}$ and $\rho_{34\vert k}$
are separable $\forall\lambda$. Moreover, as in the previous example,
the post-measurement states for different outcomes are LU equivalent
and so we only need to find $\mathcal{I}_{14}^{f}$, $\mathcal{I}_{12}^{f}$,
and $\mathcal{I}_{34}^{f}$ for a particular outcome. 

From the expressions \eqref{rho14-rank-2}, \eqref{rho12/34-rank2}
of the post-measurement states and using \eqref{I(rho)=00003DM(rho)},
we find that 
\begin{eqnarray*}
\mathcal{I}_{14}^{f} & = & 1+\left(2\lambda-1\right)^{2},\\
\mathcal{I}_{m\,m+1}^{f} & = & 4\lambda\left(1-\lambda\right),\;m=1,3
\end{eqnarray*}
for all $\lambda\in\left[0,1\right]$. Consequently, 
\begin{eqnarray*}
\mathcal{I}_{m\,m+1}^{f}+\mathcal{I}_{14}^{f} & = & 2,\;m=1,3.
\end{eqnarray*}
Hence, \emph{information} is conserved for all $\lambda\in\left[0,1\right]$.
Note however that entanglement is not conserved in any bipartition
$\forall\lambda\in\left(0,1\right)$ because (a) the states $\rho_{m\,m+1}$,
$m=1,3$ are separable and (b) $\rho_{14}$ is not maximally entangled
whenever $\lambda\in\left(0,1\right)$. 

Interestingly, \emph{information} is conserved even at $\lambda=\frac{1}{2}$,
i.e., when not only the measurement is separable but also all three
pairs become separable. So separable measurements can conserve \emph{information}
even when all post-measurement states turn separable after Bob's measurement.
The functions $\mathcal{I}_{14}^{f}$, $\mathcal{I}_{12}^{f}$, and
$\mathcal{I}_{{\rm tot}}^{f}$ are plotted in Fig.$\,$2, where $\mathcal{I}_{{\rm tot}}^{f}=\mathcal{I}_{12}^{f}+\mathcal{I}_{14}^{f}$.
As one can see, \emph{information} gained in $\left(1,4\right)$ is
always the same as \emph{information }lost from the pair $\left(1,2\right)$
for all $\lambda$. 

\begin{figure}
\includegraphics[scale=0.4]{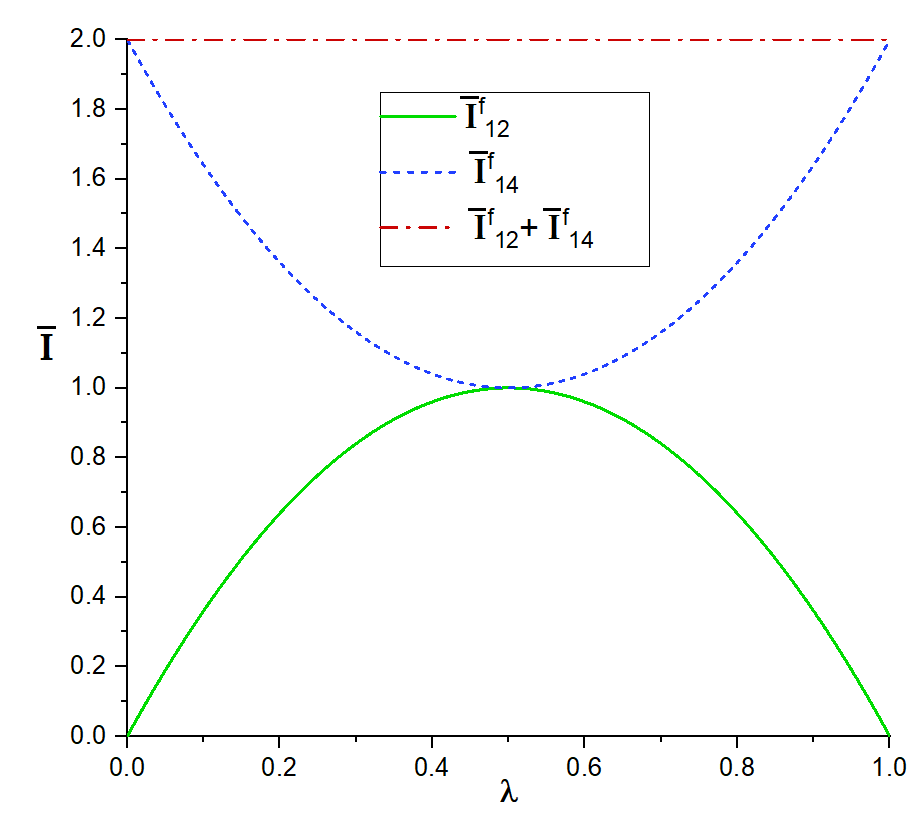}\caption{Rank-2 Bell-diagonal measurement: Behaviour of $\mathcal{I}_{14}^{f}$
, $\mathcal{I}_{12}^{f}$, and $\mathcal{I}_{{\rm tot}}^{f}=\mathcal{I}_{12}^{f}+\mathcal{I}_{14}^{f}$
as function of $\lambda\in\left[0,1\right]$.}
\end{figure}

\section{Conclusions}

In this paper, we investigated information-disturbance trade-off in
generalized entanglement swapping protocols. In particular, we considered
protocols where starting from two Bell pairs $\left(1,2\right)$,
$\left(3,4\right)$ shared between Alice and Bob, Bob and Charlie,
respectively, Bob performs a two-qubit measurement on $\left(2,3\right)$
and communicates the outcome to Alice and Charlie, so that $\left(1,4\right)$
becomes correlated. We obtained trade-off inequalities between \emph{information}
gain in $\left(1,4\right)$ and residual \emph{information} in the
pairs $\left(1,2\right)$ and $\left(3,4\right)$, respectively, where
\emph{information} is understood as the total information contained
in correlations of a two-qubit state, which is quantified in terms
of an information theoretic measure of entanglement. We argued that
when equality holds for any given measurement then it implies \emph{information}
is conserved across the bipartitions $A\vert BC$ and $C\vert BA$.
We further studied these inequalities for some well-known two-qubit
measurements and found the inequalities to be strict for complete
orthogonal but nonmaximally entangled measurements and for Bell measurement
mixed with white noise. However, rather counter-intuitively, we found
that there exist rank-2 Bell-diagonal measurements that conserve \emph{information}
but do not conserve entanglement, and moreover, there exist separable
measurements that conserve information even when all three pairs $\left(1,2\right)$,
$\left(3,4\right)$, and $\left(1,4\right)$ become separable after
Bob's measurement. This is particularly interesting because it shows
that correlations in an entangled pair can be distributed in separable
pairs in nontrivial ways so that there is no loss of \emph{information}.

Our results open up some interesting avenues for further research.
Let us begin by noting that the inequalities \eqref{trade-offs-avg-2}
are basic in nature as they hold for all entanglement swapping protocols.
On the other hand, the second set of inequalities given by \eqref{trade-offs-avg-2-1},
and the subsequent interpretation of conservation/loss of \emph{information},
crucially depends on the fact that the initial states are maximally
entangled. So it would be very interesting to obtain inequalities
similar to that of \eqref{trade-offs-avg-2-1} for arbitrary initial
states. This might, however, require considering a different \emph{information}
measure, which should ideally be a LOCC monotone. 

Finally, we would like to remark that it is important to explore and
understand the basic questions related to information transfer in
entanglement swapping-like protocols, and we hope our results would
stimulate further research on this topic. 
\begin{acknowledgement*}
S.B. is supported in part by SERB (Science and Engineering Research
Board), Department of Science and Technology, Government of India
through Project No. EMR/2015/002373. 
\end{acknowledgement*}


\begin{thebibliography}{99}
\bibitem{swapping} M. Zukowski, A. Zeilinger, M. A. Horne, and A.
K. Ekert, `{}`Event-ready-detectors'{}' Bell experiment via entanglement
swapping, \href{https://doi.org/10.1103/PhysRevLett.71.4287}{Phys.
Rev. Lett. \textbf{71}, 4287 (1993)}.

\bibitem{Bose-1998-swapping-multiparty} S. Bose, V. Vedral, and P.
L. Knight, Multiparticle generalization of entanglement swapping,
\href{https://doi.org/10.1103/PhysRevA.57.822}{Phys. Rev.A \textbf{57},
822 (1998)}.

\bibitem{Gour-Sanders-2004} G. Gour and Barry C. Sanders, Remote
preparation and distribution of bipartite entangled states, \href{https://doi.org/10.1103/PhysRevLett.93.260501}{Phys.
Rev. Lett. \textbf{93}, 260501 (2004)}.

\bibitem{Perseguers-2008-networks} S. Perseguers, C. I. Cirac, A.
Acin, M. Lewenstein, and J. Wehr, Entanglement distribution in pure-state
quantum networks, \href{https://doi.org/10.1103/PhysRevA.77.022308}{Phys.
Rev.A \textbf{77}, 022308 (2008)}.

\bibitem{networks-review} S. Perseguers, G. J. Lapeyre Jr, D. Cavalcanti,
M. Lewenstein, and A. Acin, Distribution of entanglement in large-scale
quantum networks, \href{https://doi.org/10.1088/0034-4885/76/9/096001}{Rep.
Prog. Phys. \textbf{76}, 096001 (2013)}.

\bibitem{swapping-CHSH} A. Wojcik, J. Modlawska, A. Grudka and M.
Czechlewski, Violation of Clauser-Horne-Shimony-Holt inequality for
states resulting from entanglement swapping, \href{https://doi.org/10.1016/j.physleta.2010.09.069}{Phys.
Lett. A \textbf{374}, 48, 4831 (2010)}.

\bibitem{swapping-activation} W. Klobus, W. Laskowski, M. Markiewicz,
and A. Grudka, Nonlocality activation in entanglement-swapping chains,
\href{https://doi.org/10.1103/PhysRevA.86.020302}{Phys. Rev. A \textbf{86},
020302(R) (2012)}.

\bibitem{Zeilinger-99} A. Zeilinger, A foundational principle for
quantum mechanics, \href{https://doi.org/10.1023/A:1018820410908}{Foundations
of Physics \textbf{29}, 631-643 (1999)}.

\bibitem{BZ-1999} C. Brukner and A. Zeilinger, Operationally Invariant
Information in Quantum Measurements, \href{https://doi.org/10.1103/PhysRevLett.83.3354}{Phys.
Rev. Lett. \textbf{83}, 3354 (1999)}.

\bibitem{BZZ-2001} C. Brukner, M. Zukowski, and A. Zeilinger, The
essence of entanglement, \href{https://arxiv.org/abs/quant-ph/0106119}{quant-ph/0106119}.

\bibitem{complementarity} C. Brukner, M. Aspelmeyer, and A. Zeilinger,
Complementarity and information in delayed-choice for entanglement
swapping, \href{https://dx.doi.org/10.1007/s10701-005-7355-2}{Foundations
of Physics \textbf{35}, 11 (2005)}.

\bibitem{Camlet-2017} S. Camalet, Measure-independent anomaly of
nonlocality,\href{https://journals.aps.org/pra/abstract/10.1103/PhysRevA.96.052332}  {Phys. Rev. A {\bf{96}}, 052332 (2017)}. 

\bibitem{I-not-monotone} That $\mathcal{I}$ is a not a LOCC monotone
can be understood as follows. Recall that, for any two-qubit state,
$\mathcal{I}\left(\rho\right)=\frac{B\left(\rho\right)^{2}}{4}$,
where $\mathcal{B}\left(\rho\right)$ is the maximal mean value of
the Bell-CHSH observable. Recently, it has been proven that any Bell
local $\rho$ with hidden nonlocality can be deterministically transformed
into a nonlocal state using LOCC \citep{Camlet-2017}. So $\mathcal{B}\left(\rho\right)$
is not a LOCC monotone and therefore neither is $\mathcal{I}\left(\rho\right)$. 

\bibitem{Bell-monogamy} B. Toner and F. Verstraete, Monogamy of Bell
correlations and Tsirelson's bound, \href{https://arxiv.org/abs/quant-ph/0611001}{arXiv:quant-ph/0611001}.

\bibitem{Bell-monogamy-2} S. Cheng and M. J. W. Hall, Anisotropic
invariance and the distribution of quantum correlations, \href{https://doi.org/10.1103/PhysRevLett.118.010401}{Phys.
Rev. Lett. \textbf{118}, 010401 (2017)}.

\bibitem{Horodecki-1995-Bell} R. Horodecki, P. Horodecki, and M.
Horodecki, Violating Bell inequality by mixed spin-1/2 states: necessary
and sufficient condition, \href{https://doi.org/10.1103/Phys.Lett.A.200.340}{Phys.Lett.A
\textbf{200},340 (1995)}.

\bibitem{Horodecki-1996-Bell-2} R. Horodecki, Two spin-1/2 mixture
and Bell inequalities, \href{https://doi.org/10.1103/Phys.Lett.A.210.223}{Phys.Lett.A
\textbf{210},223 (1996)}.
\end{thebibliography}
\end{document}